\pdfoutput=1
\documentclass[conference]{IEEEtran}

\IEEEoverridecommandlockouts
\usepackage[nocompress]{cite}
\usepackage{amsmath,amssymb,amsfonts}
\usepackage[hidelinks]{hyperref}
\usepackage[capitalise]{cleveref}
\usepackage{comment}
\usepackage{booktabs}
\usepackage{multirow}
\usepackage{graphicx}
\usepackage{pgfplots,pgfplotstable}

\usepackage{subcaption}
\usepackage{textcomp}
\usepackage{xcolor}
\usepackage{svg}
\usepackage{xurl}
\usepackage{microtype} 
\usepackage[inline,shortlabels]{enumitem}
\def\BibTeX{{\rm B\kern-.05em{\sc i\kern-.025em b}\kern-.08em
    T\kern-.1667em\lower.7ex\hbox{E}\kern-.125emX}}
\pgfplotsset{compat=1.16}
\usepgfplotslibrary{colorbrewer}

\usepackage[eulergreek]{sansmath}
\usepackage{filecontents}

\pgfplotsset{
    tick label style = {font=\sansmath\sffamily\small},
    every axis label = {font=\sansmath\sffamily\small},
    legend style = {font=\sansmath\sffamily\small},
    label style = {font=\sansmath\sffamily\small},
    box plot/.style={
        /pgfplots/.cd,
        black,
        only marks,
        mark=-,
        mark size=\pgfkeysvalueof{/pgfplots/box plot width},
        /pgfplots/error bars/y dir=plus,
        /pgfplots/error bars/y explicit,
        /pgfplots/table/x index=\pgfkeysvalueof{/pgfplots/box plot x index},
    },
    box plot box/.style={
        /pgfplots/error bars/draw error bar/.code 2 args={%
            \draw  ##1 -- ++(\pgfkeysvalueof{/pgfplots/box plot width},0pt) |- ##2 -- ++(-\pgfkeysvalueof{/pgfplots/box plot width},0pt) |- ##1 -- cycle;
        },
        /pgfplots/table/.cd,
        y index=\pgfkeysvalueof{/pgfplots/box plot box top index},
        y error expr={
            \thisrowno{\pgfkeysvalueof{/pgfplots/box plot box bottom index}}
            - \thisrowno{\pgfkeysvalueof{/pgfplots/box plot box top index}}
        },
        /pgfplots/box plot
    },
    box plot top whisker/.style={
        /pgfplots/error bars/draw error bar/.code 2 args={%
            \pgfkeysgetvalue{/pgfplots/error bars/error mark}%
            {\pgfplotserrorbarsmark}%
            \pgfkeysgetvalue{/pgfplots/error bars/error mark options}%
            {\pgfplotserrorbarsmarkopts}%
            \path ##1 -- ##2;
        },
        /pgfplots/table/.cd,
        y index=\pgfkeysvalueof{/pgfplots/box plot whisker top index},
        y error expr={
            \thisrowno{\pgfkeysvalueof{/pgfplots/box plot box top index}}
            - \thisrowno{\pgfkeysvalueof{/pgfplots/box plot whisker top index}}
        },
        /pgfplots/box plot
    },
    box plot bottom whisker/.style={
        /pgfplots/error bars/draw error bar/.code 2 args={%
            \pgfkeysgetvalue{/pgfplots/error bars/error mark}%
            {\pgfplotserrorbarsmark}%
            \pgfkeysgetvalue{/pgfplots/error bars/error mark options}%
            {\pgfplotserrorbarsmarkopts}%
            \path ##1 -- ##2;
        },
        /pgfplots/table/.cd,
        y index=\pgfkeysvalueof{/pgfplots/box plot whisker bottom index},
        y error expr={
            \thisrowno{\pgfkeysvalueof{/pgfplots/box plot box bottom index}}
            - \thisrowno{\pgfkeysvalueof{/pgfplots/box plot whisker bottom index}}
        },
        /pgfplots/box plot
    },
    box plot median/.style={
        /pgfplots/box plot,
        /pgfplots/table/y index=\pgfkeysvalueof{/pgfplots/box plot median index}
    },
    box plot width/.initial=1em,
    box plot x index/.initial=0,
    box plot median index/.initial=1,
    box plot box top index/.initial=2,
    box plot box bottom index/.initial=3,
    box plot whisker top index/.initial=4,
    box plot whisker bottom index/.initial=5,
}

\newcommand{\boxplot}[2][]{
    \addplot [box plot median,#1] table {#2};
    \addplot [forget plot, box plot box,#1] table {#2};
    \addplot [forget plot, box plot top whisker,#1] table {#2};
    \addplot [forget plot, box plot bottom whisker,#1] table {#2};
} 

\usepackage{fancyhdr} 
\usepackage{thomascustom}

\makeatletter
\renewcommand{\ALG@beginalgorithmic}{\footnotesize}
\makeatother

\ifthenelse{\boolean{anonymousMode}}{
    \renewcommand{\thanks}[1]{}
}{
}

\fancypagestyle{fancycopyright}{%
    \fancyhf{}%
    \fancyhead[C]{Author copy of paper published at \emph{2024 IEEE/ACM Symposium on Edge Computing (SEC)}\\
    \copyright2024 IEEE 
    }%
    \fancyfoot[C]{\thepage}%
}

\begin{document}


\title{HyperDrive: Scheduling Serverless Functions in the Edge-Cloud-Space 3D Continuum\\
    \thanks{*The authors Cynthia Marcelino and Thomas Pusztai have contributed equally to this work.}
}

\ifthenelse{\boolean{anonymousMode}}{
    \author{\IEEEauthorblockN{Anonymous Authors}
    \IEEEauthorblockA{}
    }
}{
    \author{
        \IEEEauthorblockN{Thomas Pusztai*}
        \IEEEauthorblockA{\textit{Distributed Systems Group, TU Wien} \\
            t.pusztai@dsg.tuwien.ac.at
        }
        \and
        \IEEEauthorblockN{Cynthia Marcelino*}
        \IEEEauthorblockA{\textit{Distributed Systems Group, TU Wien} \\
            c.marcelino@dsg.tuwien.ac.at
        }
        \and
        \IEEEauthorblockN{Stefan Nastic}
        \IEEEauthorblockA{\textit{Distributed Systems Group, TU Wien} \\
            snastic@dsg.tuwien.ac.at}
    }
}


\newacronym{IoT}{IoT}{Internet of Things}
\newacronym{LEO}{LEO}{low earth orbit}
\newacronym{EO}{EO}{earth observation}
\newacronym{ISL}{ISL}{inter-satellite laser link}
\newacronym{MCDM}{MCDM}{multi-criteria decision making}
\newacronym{DAG}{DAG}{directed acyclic graph}
\newacronym{QoS}{QoS}{quality of service}
\newacronym{SLO}{SLO}{Service Level Objective}
\newacronym{E2E}{E2E}{end-to-end}

\newcommand{\LEO}[0]{\gls{LEO}}
\newcommand{\EO}[0]{\gls{EO}}
\newcommand{\ISL}[0]{\gls{ISL}}
\newcommand{\ISLs}[0]{\glspl{ISL}}
\newcommand{\SLO}[0]{\gls{SLO}}
\newcommand{\SLOs}[0]{\glspl{SLO}}

\ifthenelse{\boolean{anonymousMode}}{
    \newcommand{\PolarisShortName}[0]{Starfield}
    \newcommand{\PolarisLongProjectName}[0]{``Name omitted for double-blind review''}
    \newcommand{\CentaurusProjectAndUrl}[0]{Cloudy Edge project\footnote{Project name changed and URL omitted for double-blind review}}
    
    \newcommand{\citeSlocVision}[0]{\cite{AnonymizedForDoubleBlindReview}}
}{
    \newcommand{\PolarisShortName}[0]{Polaris}
    \newcommand{\PolarisLongProjectName}[0]{Polaris SLO Cloud}
    \newcommand{\CentaurusProjectAndUrl}[0]{Centaurus project\footnote{\url{https://www.centaurusinfra.io}}}
    
    \newcommand{\citeSlocVision}[0]{\cite{SlocVision2020}}
}

\newcommand{\Polaris}[0]{\PolarisShortName{}}
\newcommand{\SLOC}[0]{\PolarisShortName{}}

\newcommand{\PolarisProjectUrl}[0]{\AnonymizableUrl{https://polaris-slo-cloud.github.io}}
\newcommand{\HyperDriveSchedulerUrl}[0]{\AnonymizableUrl{https://github.com/polaris-slo-cloud/hyper-drive}}

\newcommand{\StarlinkNodesCount}[0]{6,192} 
\newcommand{\ISLBandwidth}[0]{20~Gbps}
\newcommand{\SatToGroundBandwidth}[0]{5~Gbps}

\maketitle


\thispagestyle{fancycopyright}
\pagestyle{fancy}

\begin{abstract}
The number of Low Earth Orbit~(LEO) satellites has grown enormously in the past years.
Their abundance and low orbits allow for low latency communication with a satellite almost anywhere on Earth, and high-speed inter-satellite laser links~(ISLs) enable a quick exchange of large amounts of data among satellites.
As the computational capabilities of LEO satellites grow, they are becoming eligible as general-purpose compute nodes.
In the 3D continuum, which combines Cloud and Edge nodes on Earth and satellites in space into a seamless computing fabric, workloads can be executed on any of the aforementioned compute nodes, depending on where it is most beneficial.
However, scheduling on LEO satellites moving at approx. 27,000~km/h requires picking the satellite with the lowest latency to all data sources (ground and, possibly, earth observation satellites).
Dissipating heat from onboard hardware is challenging when facing the sun and workloads must not drain the satellite's batteries.
These factors make meeting SLOs more challenging than in the Edge-Cloud continuum, i.e.,  on Earth alone.
We present HyperDrive, an SLO-aware scheduler for serverless functions specifically designed for the 3D continuum.
It places functions on Cloud, Edge, or Space compute nodes, based on their availability and ability to meet the SLO requirements of the workflow.
We evaluate HyperDrive using a wildfire disaster response use case with high Earth Observation data processing requirements and stringent SLOs, showing that it enables the design and execution of such next-generation 3D scenarios with 71\% lower network latency than the best baseline scheduler.
\end{abstract}



\section{Introduction}
\label{sec:Intro}



As of 2024, there are over 8,000 low Earth orbit (LEO) satellites orbiting the Earth~\cite{LEO_SAT_stats}.
Satellites have traditionally communicated with each other via ground stations. Lately, \ISLs{} aim to connect satellites and create a large orbital network topology~\cite{NetworkTopology2020}.
Starlink is currently the largest LEO mega-constellation with about 7,000~satellites in orbit~\cite{FCC_StarlinkCompetition2024} and almost 12,000~total satellites approved by the FCC, which must be launched by 2028~\cite{StarlinkLowerOrbitAuth2019}.
By 2029 a second LEO mega-constellation is planned to be available with 3,236~satellites~\cite{FCC_KuiperAuthorization2020} and more competition is solicited by the FCC~\cite{FCC_StarlinkCompetition2024}.
ISL capability allows LEO satellites to act as ground edge nodes, processing data directly in orbit and near the data source, such as Earth Observation (EO) satellite data.
This opens up opportunities for new computing paradigms in space, such as Serverless Computing. 

Serverless Computing provides elastic scaling and infrastructure management. Serverless platforms automatically deploy functions, scaling up or down based on demand, thus avoiding idle resources~\cite{SlocVision2020}.  
To address the environmental heterogeneity of the Edge-Cloud-Space Continuum, Serverless platforms need scheduling mechanisms that identify environmental properties and their current conditions to deploy functions and meet their requirements~\cite{SCF_2022}.

Most common scheduling approaches focus on meeting requirements based on resources, network, application and energy~\cite{HowToPlace2019,SchedulingMechanisms2023}. 

\paragraph{Resource-Aware} Schedulers~\cite{K8sSchedulingFramework} ensure that functions are executed on nodes capable of handling their computational requirements to prevent overloading any single node, which could lead to performance degradation or failures. 
In the Edge Cloud Space Continuum, resource-aware scheduling mechanisms~\cite{SkippyScheduler,AdaptiveScalingK8sPods2020} dynamically allocate functions considering the infrastructure-specific resource characteristics such as CPU capacity and architecture, memory, and GPU. In Orbital Edge Computing (OEC), scheduling mechanisms~\cite{OEC_JointScheduling2024,OrbitaEdgeOffloading2022} address specific orbit characteristics such as satellite infrastructure resource and energy costs to transfer the data between satellites or to the ground stations. 
However, current approaches do not consider all aspects of Edge,  Cloud, and Space as a unified continuum. They neglect the impact of resource temperature and heat generated by the task execution. Due to the substantial temperature variations on satellites between the daylight and eclipse periods of an orbit, tasks that require intense computation can produce too much heat, putting satellite components at risk of damage from overheating~\cite{Vision6G2024,AutonomousScheduling2018}.

\paragraph{Network-Aware} Nodes at the edge typically have different network characteristics than cloud nodes. These network characteristics include variations in end-to-end latency, bandwidth availability, and link reliability~\cite{HowToPlace2019}. Network-aware schedulers~\cite{SkippyScheduler,QosAware2017} consider these characteristics to optimize function placement, ensuring efficient and reliable communication. In OEC,  schedulers~\cite{OEC_JointScheduling2024} typically also address the intermittent ISL communication between satellites and high latency communication with ground stations. However, existing OEC schedulers are not built for serverless functions, so they cannot guarantee the complete execution of serverless workflows across the Edge Cloud Space Continuum. Existing schedulers do not consider the positions of satellites, which is essential to ensure the seamless execution of serverless workflows from orbit to the Edge and Cloud. Therefore, existing schedulers fail to ensure that serverless functions can start, complete, and transfer all required data within the connectivity range of the satellite network.

\paragraph{Application and SLO-Aware} Applications have \SLOs{} that define the expected performance and availability during their execution. To meet these requirements, SLO-aware schedulers~\cite{PolarisScheduler,VelaScheduler2023} need to consider not only infrastructure properties such as resource availability, but also workload characteristics. Although OEC schedulers ensure functions can execute in a specific node, they do not guarantee workload requirements, i.e., \SLOs{}, such as maximum latency.

\paragraph{Energy-Aware} Schedulers consider the current power source and estimated task power consumption during the placement process. Energy-aware scheduling~\cite{sched_energy,EAIS2022} is crucial to prevent battery-powered devices from running out of power and to reduce overall power usage. By optimizing energy usage, schedulers ensure prolonged operational lifespans for edge devices and enhance sustainability, thus optimizing performance and longevity in the Edge Cloud Continuum. In OEC, energy-aware schedulers~\cite{OEC_JointScheduling2024,EnergySatOffloading2024} consider also the energy necessary to transmit the data either to other satellite nodes or to the ground station. However, existing schedulers overlook the satellite position during the energy consumption estimation. Despite tasks requiring a certain amount of power, the satellite can auto-recharge its batteries during the daylight periods.

Although current Serverless scheduling approaches address the heterogeneous devices on the Edge, they are not suitable for the specific environmental properties of the Edge Cloud Space 3D Continuum, such as satellite position and heat generation. Moreover, the current orbital scheduling approaches lack integration across the Edge Cloud and Space environment, essential for latency and function execution across the 3D Continuum.

In this paper, we introduce \emph{HyperDrive}, a novel Serverless platform that seamlessly integrates Edge, Cloud, and Space Computing, creating a 3D Continuum.
\ifthenelse{\boolean{anonymousMode}}{
    HyperDrive is part of ``\Polaris{}'', a SIG of a large Linux Foundation project (Names and URLs omitted for double-blind review) that provides a novel open-source platform for building unified and highly scalable public or private distributed Cloud and Edge systems and which is now expanding into the 3D Continuum.
}{
    HyperDrive is part of \Polaris{}\footnote{\PolarisProjectUrl{}}, a SIG of the Linux Foundation \CentaurusProjectAndUrl{}, a novel open-source platform for building unified and highly scalable public or private distributed Cloud and Edge systems, which is now expanding into the 3D Continuum.
}
HyperDrive leverages the specific capabilities of each layer of the 3D Continuum, such as Edge proximity to the data and satellite proximity to Earth observation data, to enable optimized serverless function deployment and execution.

The main contributions of this paper include:

\begin{enumerate}
    \item The \emph{architecture of the HyperDrive Serverless Platform}, which introduces novel components and mechanisms tailored to the unique characteristics of the 3D Continuum.
    HyperDrive enables functions to be seamlessly executed anywhere in the 3D Continuum, optimizing performance and reliability by ensuring that workflow \SLOs{} are met.

    \item The \emph{HyperDrive scheduling model} is the foundation of our Serverless platform's scheduler, which is the main focus of our paper.
    The HyperDrive scheduling model considers constraints such as resource capacity, application SLO requirements, satellite temperature, and network load to minimize the end-to-end Serverless workflow latency.

    \item Our \emph{Heuristic Scheduling Algorithms for the 3D Continuum} enable the realization of the HyperDrive scheduling model using a flexible \gls{MCDM} approach.
    It first filters out nodes that are not capable of hosting a function and, then, scores the remaining nodes according to multiple criteria to find the best suited node for a function.
    Our prototype implementation is available as open-source\footnote{\HyperDriveSchedulerUrl{}}.
    HyperDrive achieves 71\% lower \gls{E2E} network latency than the next best baseline approach.
\end{enumerate}


This paper has eight sections. 
\cref{sec:Motivation} presents the illustrative scenario and research challenges.
\cref{sec:Architecture} shows an overview of the HyperDrive Architecture for a Serverless Platform in the 3D Continuum.
\cref{sec:Scheduling} describes the Serverless Workflow Model, HyperDrive scheduling optimization model, and heuristic scheduling algorithms for the 3D Continuum.
\cref{sec:ImplementationAndExperimentsDesign} details our implementation approach and describes the design of our experiments.
\cref{sec:Results} discusses the results of the experiments, 
\cref{sec:Relatedwork} presents related work. 
\cref{sec:Conclusion} concludes the paper and outlines our future work.

\section{Motivation}
\label{sec:Motivation}

To further motivate our work we present an illustrative disaster response scenario and leverage it to derive research challenges.

\subsection{Illustrative Scenario}

\begin{figure}[tb]
\centering
\includegraphics[width=.99\linewidth]{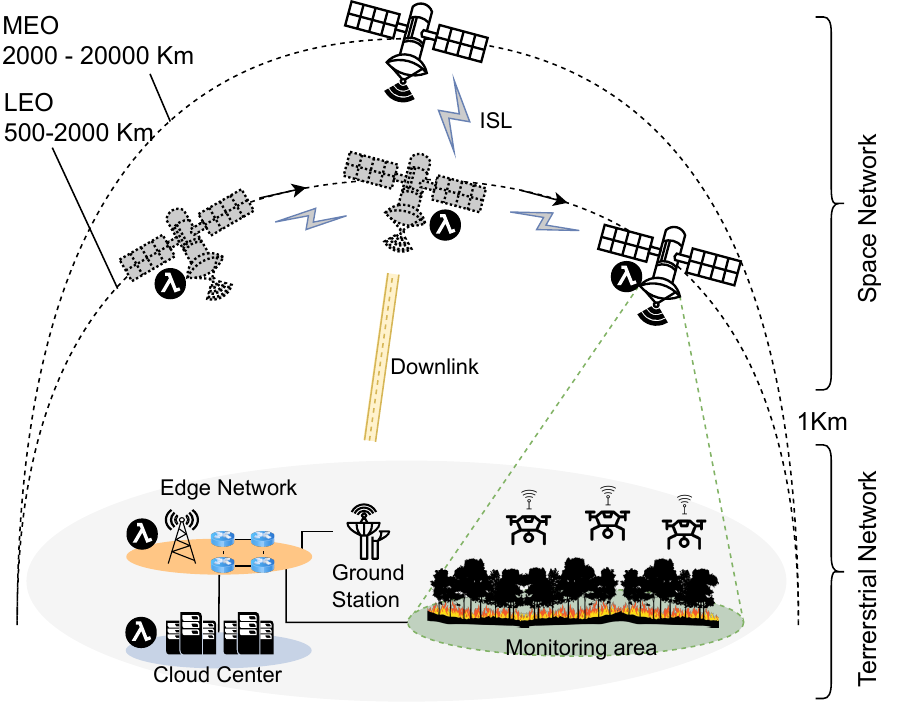}
\caption{Illustrative Scenario: Wildfire detection with on-ground and in-orbit Serverless Edge Computing}
\label{fig:IllustrativeScenario}
\end{figure}

Early detection of wildfires in remote areas is critical to mitigate their effects. Our scenario (\cref{fig:IllustrativeScenario}) involves using a combination of drones, LEO satellites, and ground-based Edge nodes that compose a serverless workflow for real-time wildfire detection, inspired by~\cite{InteagratingEdgeLEO2021,SkyEdge2021,InorbitML2023,FireSatBlog2024}. The drones operate in high-risk wildfire areas, such as California during the summer, monitoring specific zones and capturing video and sensor data to watch for signs of wildfires. They send the data to the nearest Edge node using streaming frameworks or, when out of range, transmit it to LEO satellites acting as in-orbit Edge nodes. Once a fire is detected, LEO satellites incorporate satellite Earth Observation (EO) data for processing. Our serverless workflow processes the data close to the source to improve latency and reduce network overhead. In some situations, functions are executed directly on LEO satellites due to the data's proximity to EO data and the high latency associated with downloading data to the ground.

\begin{figure}[tb]
\centering
\includegraphics[width=.9\linewidth]{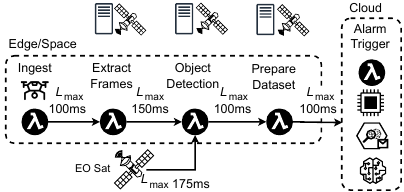}
\caption{Simplified Serverless Workflow for Wildfire Detection}
\label{fig:workflow}
\end{figure}

~\cref{fig:workflow} shows our Serverless workflow with four Serverless functions, partially executed on the Edge, partially executed in-orbit and partially executed in the Cloud. During the \textit{Ingest} stage, real-time videos are transmitted to Edge nodes on the ground or in-orbit. The \textit{Extract Frames} function processes small video chunks received from Ingest stage and extracts image frames. \textit{Object Detection} functions identify wildfire patterns in the extracted images, such as smoke, flames, or hotspots. The \textit{Prepare Dataset} function prepares the data for resource-intensive tasks. The processed data is transmitted to the Cloud for storage and more resource-intensive tasks, such as machine learning model inference. In the Cloud, \textit{Alarm trigger} functions evaluate the data and decide whether to trigger local emergency responses or deploy more drones to a specific area to confirm the wildfire before triggering an alarm.


Serverless computing allows dynamic scaling and processing close to the data source. By running Serverless functions directly on LEO satellites, we can combine data from the drones on the Earth and from EO satellites to process data as soon as they are produced.
Atmospheric interference reduces link speeds to ground stations, typical speeds are around 300~Mbps\cite{EDRS_Overview}.
Thus, downlinking data from EO satellites to Earth would take too long due to the large volume of data, e.g., each of the ESA Sentinel~2 satellites supplies high resolution images for a swath of 290~km in 13~spectral bands, producing about 1.5~TB of data per day~\cite{ESA_Sentinel2Ops, Sentinel2CLaunched2024}.
Since EO satellites only downlink to dedicated ground stations, the data may even be queued~\cite{L2D2_2021}.
For Sentinel-2 ``real-time'' product availability is defined as ``no later than 100 minutes after data sensing''~\cite{SentinelGlossary}, which violates the satellite data ingestion link SLO of the wildfire application. 
\ISLs{} between EO satellites and LEO satellites are much better suited for large EO data volumes, since their speeds can be much higher -- recently a 100~Gbps \ISL{} from GEO to LEO has been demonstrated~\cite{Geo2LeoSat100Gbps2024}.
Hence, it is much faster to uplink a one GB ML model to the satellite than to downlink the EO data to a ground station.
Drone videos are also moderate in size, e.g., a three minute 4K video from the FLAME2 dataset~\cite{WildlandFireDet2022} amounts to 2.2~GB, which qualifies for uplinking to a LEO satellite in real-time.

Combining satellite EO data with drone data on LEO satellites allows reducing the time it takes to analyze and respond to wildfires.
Additionally, it provides a reliable alternative when Edge nodes are out of range or experiencing connectivity issues. 
Scheduling the functions to execute in orbit ensures that wildfire detection and monitoring continue uninterrupted, even if ground-based infrastructure faces limitations. It allows immediate data processing and decision-making in orbit, reducing delays and ensuring continuous, real-time monitoring. As a result, we can decrease response times to wildfire threats. However, there are several challenges associated with scheduling Serverless functions in 3D continuum.

\subsection{Research Challenges for Scheduling in the 3D Continuum}\label{MotivationRequirements}
Based on the illustrative scenario, we identify several key requirements for scheduling serverless functions on LEO satellites in orbit as follows:
\begin{researchChallenges}
    \rc{
        \emph{Satellite Availability}: Unlike Edge nodes, which have fixed positions, LEO satellites are constantly in motion as they orbit the Earth, which impacts their availability and communication windows~\cite{GearingUp2018}. A satellite must be within range of \textcircled{\footnotesize{1}} the drone, \textcircled{\footnotesize{2}} the EO satellite, and \textcircled{\footnotesize{3}} the ground station to be considered available for scheduling. Specifically, the satellite needs to be within the drone's range to receive real-time video transmissions from Earth. At the same time, it must also be within the range of the EO satellite to receive and relay additional monitoring data. In addition, the satellite must be within range of ground stations, which have Cloud control planes for tasks such as scheduling. However, the term ``in range'' is more complex than direct line of sight. Since satellites can communicate via \ISLs{}~\cite{DelayNotAnOption2018, AnalysisISL2021}, a satellite can be in range, if the bandwidth and latency via \ISLs{} is acceptable for the purpose of the communication (e.g., data transfer).
        According to a recent study~\cite{StarlinkPerf2024} Starlink's median roundtrip latency (client-LEO-Cloud) is 40-50~ms; the theoretical roundtrip latency between New York and London when routing exclusively through \ISLs{} is 58-66~ms~\cite{DelayNotAnOption2018}.
        As satellites move in and out of range, the Serverless platform must continuously adapt, reallocating resources and re-establishing communication links. Therefore, satellite availability is more dynamic and complex compared to static Edge nodes.
    }\label{SatPos}
    \rc{
        \emph{Power Supply}: The scheduler must consider the satellite's power state, including its batteries' current charge level and the overall health of its energy storage system. Given the increasing computing power in satellites and the strict size constraints for some of them~\cite{Vision6G2024}, the scheduler must be aware of the energy requirements of specific serverless functions to ensure that the satellite has enough power reserves to execute these functions without depleting its energy resources. Finally, a CubeSat's solar panels produce only up to 7~W of power~\cite{SmallSatellitesSurvey2018}, while batteries can have a density of up to 190~Wh/kg~\cite{NASA_SmallSpacecraftTech2024}. This means that a satellite might not be designed to fully recharge their batteries in a daylight period of an orbit. Thus, the scheduler must evaluate whether the power expenditure of its workloads can be compensated with solar power before the battery depletes.
    }\label{Power}
    \rc{
        \emph{Computing Capacity \& Heat Generation}: LEO satellites are deployed with fixed and limited resources that cannot be patched or upgraded throughout their lifetime. These satellites are built to consume minimal energy and are equipped with minimal components to reduce weight and, consequently, launch costs. As computing increases, the temperature also rises. Since there is no atmosphere in space, heat dissipation mainly occurs through thermal radiation and lack of exposure to the sun. LEO satellites typically face temperatures from $-120^{\circ}$C in the shade to $+120^{\circ}$C when in the sunlight~\cite{SpaceEnvEffectsNASA2020}.  This situation can lead to prolonged high temperatures, affecting the performance of critical components such as the CPU~\cite{Vision6G2024,SCCaseStudy2023,SatelliteComputingVision2023,SkyEdge2021}. Therefore, the scheduler must consider not only the existing processing capacity but also the current temperature of the components and how long they potentially need to dissipate the heat.
    }\label{Resources}
    \rc{
        \emph{Scalability}: Due to the fixed number of satellites in orbit and the increased costs associated with launching new ones, horizontal scaling presents a significant challenge. Compared to the ground data centers, where additional servers can be easily deployed to meet increasing demand, the satellite network is limited by the number of satellites currently in orbit. This physical resource constraint and fixed number of nodes make it challenging to auto-scale effectively to meet varying workload demands~\cite{LeoComputingPlatform2021,ServerlessAbstraction2023}.
    }\label{r:3}
    \rc{
        \emph{SLO Awareness}: Serverless workflows must meet specific Service Level Objectives (SLOs) to ensure performance and reliability. These \SLOs{} typically include minimal latency and bandwidth, which are essential for maintaining optimal service performance. Maintaining \SLOs{} on the ground can already be challenging~\cite{PolarisMiddleware2021,SloScript2021} and these challenges are exacerbated by the network specifics, orbital movements, battery, and heat conditions of satellites~\cite{QoSLeoPlacement2022}. Therefore, to ensure performance and reliability, the scheduler must consider the state of multiple nodes when enforcing workload \SLOs{}.
    }\label{SloAwareness}
  \rc{
        \emph{Workflow Dependencies}: In a mixed environment, Serverless workflows can be executed on ground-based or LEO Edge nodes. The scheduler needs to take into account the workflow composition to identify the dependencies and interactions between the functions. Additionally, the scheduler must consider the placement of these functions within the workflow to ensure that interdependent tasks are located closely together to minimize latency and maximize efficiency~\cite{SkippyScheduler}
    }\label{workflowDependency}
    
\end{researchChallenges}

\section{Architecture Overview of a Serverless Platform for the 3D Continuum}
\label{sec:Architecture}

\begin{figure*}[!htbp]
\centering
\includegraphics[width=.8\linewidth]{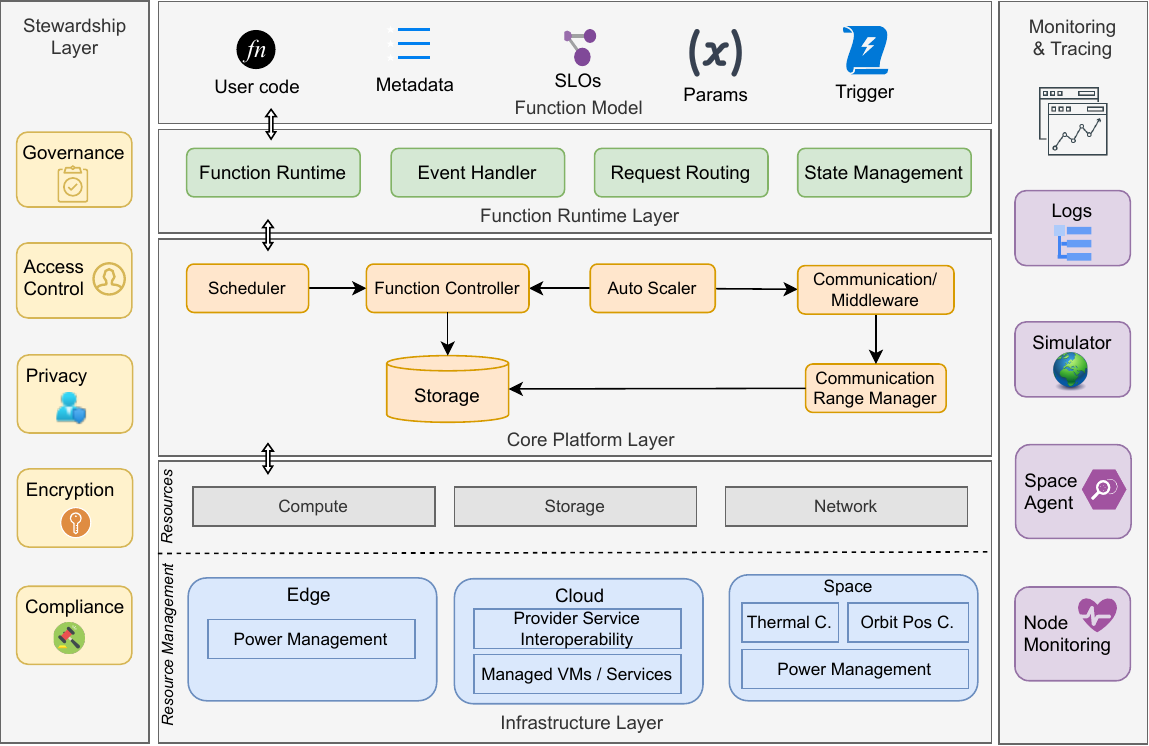}
\caption{Architecture Overview of a Serverless Platform for the Edge-Cloud-Space 3D Continuum}
\label{fig:architecture}
\end{figure*}

HyperDrive is a novel serverless platform specifically designed for the 3D Continuum, as shown in ~\cref{fig:architecture}. To achieve that, our platform proposes six different layers:  
\begin{enumerate*} [(a)]
\item an infrastructure layer that unifies the computing resources in the 3D Continuum,
\item a core platform layer for efficient and optimized function deployment and execution,
\item a function runtime layer for lightweight and low-latency execution,
\item a function model to allow developers change function behavior,
\item monitoring and tracing for real-time insights and
\item a stewardship layer layer composed of frameworks that enforce governance, security and compliance.
\end{enumerate*}
Each platform layer introduces components to address the research challenges presented in \cref{MotivationRequirements}. 

\subsection{Infrastructure Layer}
This layer includes common computing resources across the Edge-Cloud-Space 3D Continuum, such as computing, storage, and network. Each computing layer, i.e., Edge, Cloud, and Space within the 3D Continuum, has specific properties that require tailored resource management. In the Edge layer, the HyperDrive infrastructure layer manages battery power to prevent Edge devices from running out of power. For example, by providing battery level information to the scheduler so that only drones with enough battery capacity execute the \inlinecode{Ingest} function in the wildfire serverless workflow. In the Cloud, it handles heterogeneous provider-managed services such as AWS S3 storage and Azure storage for storing high-resolution satellite images or more intense computing tasks such as running inference on machine learning models. In the space layer, the platform manages thermal regulation and power to prevent satellite depletion. Furthermore, the infrastructure layer provides satellite positioning information, which is critical for HyperDrive scheduler to place functions within range to ensure efficient data exchange between the functions.
These computing resources create a unified infrastructure layer that adapts to the heterogeneous and dynamic requirements of the 3D Continuum, enabling the HyperDrive Serverless Platform to adjust to resources based on demand, ensuring seamless execution across the Edge-Cloud-Space Continuum. This is a key prerequisite to achieving our vision of self-provisioning infrastructures~\cite{nastic2024selfprovisioningInfrastructure}.

\subsection{Core Platform Layer}
This layer incorporates components responsible for managing and orchestrating tasks across the 3D Continuum. It manages the configuration, deployment, computation balancing~\cite{kexin2023AffentionFunc}, and auto-scaling of serverless functions, handling their lifecycle and scaling resources up and down based on the workload demand, such as the wildfire serverless workflow. Moreover, the storage enables HyperDrive to store function deployment properties and specific function configurations, such as parameters, state management settings, and \SLOs{}. To allocate functions effectively, HyperDrive scheduler considers the 3D Continuum requirements described in \cref{MotivationRequirements}. HyperDrive scheduler utilizes resource-based scheduling mechanisms, commonly used by various Edge and Cloud schedulers~\cite{SkippyScheduler,PolarisScheduler,VelaScheduler2023}. HyperDrive considers different requirements, including resource capacity, workload SLO, power supply, and satellite position, to make decisions using an \gls{MCDM} approach.
To ensure scalability in the large 3D Continuum, HyperDrive is a distributed scheduler that operates with multiple instances.
Distributed scheduling requires keeping node state information in sync among the scheduler instances and handling scheduling conflicts.
To address these two challenges, each HyperDrive scheduler instance obtains a function's candidate nodes and their states from the Monitoring Agent using sampling, similar to other distributed schedulers~\cite{TarcilScheduler2015, VelaScheduler2023}, and handles conflicts using the MultiBind mechanism described in \cref{sec:Scheduling.Algorithms}.
By integrating Edge-Cloud-Space requirements, HyperDrive ensures the optimal placement and performance of Serverless functions within the 3D Continuum, thus meeting application demands and respecting boundaries between ground and space requirements such as latency and financial costs.

\subsection{Function Runtime Layer}

The Function Runtime layer consists of components such as Function Runtime, Event Handler, Request Routing, and State Management. The Function Runtime relies on lightweight frameworks such as WebAssembly to provide safety, isolation, and low-latency communication~\cite{Cwasi2023,faasm2020}. In our illustrative scenario, the runtime utilizes function locality to reduce network overhead, ensuring satellites leverage local mechanisms such as inter-process communication (IPC) to exchange data between functions on the same host. Thus, the function runtime reduces latency and ensures that communication between co-located functions remains local, avoiding unnecessary ISL communication. Serverless stateless design pushes functions to leverage external services for state management~\cite{SCF_2022,WhereWeAreLiesAhead}. HyperDrive State Management leverages mechanisms such as short-term memory state~\cite{goldfish2024, goronjic2024miso} to allow serverless workflows, like wildfire detection, to maintain their state between executions, thereby avoiding the overhead of external service communication. Due to the different properties, such as bandwidth, latency, and jitter, between Edge, Cloud, and Space, HyperDrive Request Routing optimizes load balancing by forwarding requests to functions in the vicinity, thus reducing latency by avoiding communication between functions cross-environment, such as Edge and space. The Event Handler manages events from different sources, such as image drones and EO data, to ensure proper function invocation. The components in this layer ensure a seamless execution of serverless functions to meet the workload requirements effectively. The function runtime layer offers lightweight mechanisms for executing functions on limited resource devices across the 3D Continuum.

\subsection{Function Model}

This layer introduces a function model that allows developers to define specific behaviors, such as \SLOs{} and trigger types, in addition to the function code, parameters, and metadata. 
Developers can specify the type of event - such as streaming, asynchronous, or synchronous - that the function should process. In the 3D continuum, the function model enables users to react to specific satellite events, such as changes in orbit or satellite payload data received. Specifically, in the wildfire serverless workflow, drones at the Edge trigger \inlinecode{ExtractFrames} function using video streams, while \inlinecode{ObjectDetection} are triggered by single image frames as data input. Moreover, developers may specify certain \SLOs{}, such as a maximum latency of 100~ms between two functions, for instance, between \inlinecode{ExtractFrames} and \inlinecode{ObjectDetection}. Without coding effort, developers can indicate whether functions are stateless or stateful. The HyperDrive Function Model layer abstracts the underlying infrastructure, enabling developers to manage serverless workflows without the complexity of coding or infrastructure management. Finally, this layer offers specifically tailored programming modes, e.g., to facilitate dealing with large-scale, heterogeneous data sources~\cite{sehic2012ProgrammingModel}.

\subsection{Monitoring \& Tracing}
This layer is composed of components that enable real-time tracking and monitoring such as Space Agent, Node Monitoring, distributed logging systems, and a simulator that enables developers to simulate functions execution without deploying the function on the expensive and limited infrastructure, e.g., on the satellites. 
The Monitoring Agent is designed to track and analyze key performance metrics across the 3D Continuum, including Edge, Cloud, and space infrastructure. 
It watches computing capacity, memory usage, and resource utilization across all nodes to prevent overloading and ensure efficient function execution. 
Additionally, it monitors network quality of service (QoS) parameters, including bandwidth and latency, to maintain compliance with workload SLOs.
By monitoring the common properties of different layers, the Monitoring Agent enables seamless integration and reliability across the 3D Continuum.

The Space Agent is specifically designed to address the requirements of in-orbit computing. It is responsible for tracking the unique properties of the space environment, including the availability of LEO satellites, taking into account their rapid movement in orbit and their limited communication windows.
Additionally, the Space Agent manages \ISLs{} and ground-satellite network graphs to ensure that the satellite can meet the user-defined latency \SLOs{}. 
It also monitors the satellite power supply, identifying the current charge levels of batteries and their position in relation to solar energy generation, to ensure that serverless functions are assigned only to satellites with sufficient battery capacity.
Furthermore, the Space Agent monitors satellite thermal levels to prevent overheating caused by high computational load or prolonged usage, which could result in execution failures and potentially lead to long-term hardware damage. By addressing these space-specific requirements, the Space Agent plays a crucial role in optimizing the scheduling, deployment and execution of serverless functions across the 3D Continuum.

\subsection{Stewardship Layer}

This layer ensures serverless functions' secure, compliant, and efficient operation across the 3D Continuum. Its components enforce compliance with environmental and data protection regulations relevant to the workflow, such as wildfire monitoring. Encryption leverages mechanisms to protect stored sensitive information, while privacy mechanisms ensure that personal or location-based data is handled confidentially by the system. Moreover, Access Control implements role-based access and fine-grained permissions to restrict unauthorized access and actions.
At the same time, the Governance component oversees these processes, enforcing policies and standards to maintain system integrity, security, and performance across the platform under expected conditions but also under uncertainty~\cite{nastic2015governinguncertainty}.

\section{HyperDrive SLO-Aware Scheduler for 3D Continuum}
\label{sec:Scheduling}

The HyperDrive scheduler is designed to address the challenges that arise in the placement of serverless functions in the 3D Continuum first using an optimization problem and, then, using an \gls{MCDM} approach.
Without loss of generality, we assume that every serverless function is part of a serverless workflow, which we model as follows.

\subsection{Serverless Workflow Model}

A serverless workflow can be modeled as a \gls{DAG} with every node representing an executable task, i.e., a serverless function or an operator, such as a condition, fork, or loop, and every link representing an invocation of the next node.
The workflow \gls{DAG} for our wildfire detection use case is is part of \cref{fig:workflow}; all executable tasks are by nodes with a $\lambda$ sign.
For the purpose of scheduling we refer to a serverless function instance as a \emph{task}.


The workflow graph can be annotated with metadata relevant to its tasks.
Each task node is annotated with information such as container image, resource requirements, preferred location, and \SLOs{}.
Since many network connections in the 3D Continuum are not as reliable as within a Cloud data center, tasks need to be able to specify special needs regarding the network \gls{QoS} for incoming and outgoing links.
To this end each workflow link can be annotated with network \SLOs{}, specifically with maximum allowed latency, minimum bandwidth, maximum jitter, and maximum packet drop percentage.

In many cases serverless functions do not only depend on data from the predecessor function(s), but also on an external data source.
In the 3D Continuum such an external data source may be, e.g., an S3 storage in a Cloud data center or high resolution data from an EO satellite.
Workflow \SLOs{} may result in special requirements for the connections to these data sources, i.e., network \gls{QoS} \SLOs{}.
This entails that a workflow \gls{DAG} must capture not only executable nodes, but also data source nodes and support \SLOs{} on their outgoing links.
The ``EO Sat'' at the bottom of \cref{fig:workflow} represents an EO satellite node as a data source with its outgoing link providing EO data and imposing a max latency \SLO{} of 175~ms to the \inlinecode{ObjectDetection} function.
This metadata gives the HyperDrive scheduler all the required information to make a suitable placement of the workflow's tasks.

\subsection{HyperDrive Scheduling Model}

Let a Serverless workflow be a DAG \( \mathcal{W=(F,E)} \), where each node in the DAG represents a function in the Set \( \mathcal{F} \) and each edge in Set \( \mathcal{E} \) represents the invocation of the next task. Let the network graph be \( \mathcal{G=(N,L)} \), where  \( \mathcal{N} \) is a Set of nodes and  \( \mathcal{L} \) the communication latency between the nodes. The scheduling goal to minimize the latency in the Serverless workflow \( \mathcal{W} \) execution in the 3D Continuum, effectively mapping the workflow \( \mathcal{W} \) onto the network graph \( \mathcal{G} \). To achieve this, we consider the following constraints:

\emph{Resource Capacity:} This constraint ensures that every node has enough resources to process the scheduled function, maintaining system stability and performance. Additionally, this constraint helps balance the system load across the nodes, optimizing the overall utilization of available resources. Therefore, the total resource demand $D_{i}$ of function $i$ on each node $n$ in \( \mathcal{N} \) must not exceed its availability resources $R_n$:

\begin{equation}
    \sum_{i \in \mathcal{F}} D_{i} \leq R_{n} \quad \forall n \in \mathcal{N}
\end{equation}

\emph{Network SLOs:} This constraint ensures that data transfer between functions occurs within acceptable timeframes, ensuring that functions perform as expected. This means that communication between functions must meet performance criteria defined by the user to minimize delays. Thus, the SLOs latency $S_{ij}$ must be met for each function invocation pair ($i,j)$ in functions \( \mathcal{F} \). The latency $L_{nm}$ of the path between nodes $n,m$ in \( \mathcal{N} \) must not exceed the SLO $S{_ij}$:

\begin{equation}
L_{nm} \leq S_{ij} \quad \forall (i,j) \in \mathcal{F}, \forall (n,m) \in \mathcal{N}    
\end{equation}

\emph{Temperature:} Managing thermal conditions not only protects the physical integrity of the nodes but also maintains optimal performance and longevity, specially in space where extreme temperature variations are common. Therefore, the temperature of each node $n$ in \( \mathcal{N} \) must not exceed its maximum allowed temperature $T_{\text{max}}$ , considering the maximum temperature caused by the satellite exposure to the sun and the temperature sum increase due to the execution of the each function $T_{\text{exc}}$:

\begin{equation}
    T_{\text{orb}}^{n} + \sum_{i \in \mathcal{F}} T_{\text{exc}}^{in} \leq T_{\text{max}}^{n} \quad \forall n \in \mathcal{N}
\end{equation}

The scheduler goal is to minimize the total latency in the workflow execution by summing the latency $L_{nm}$ between nodes $n,m$ in \( \mathcal{N} \) for each function invocation $i,j$ in \( \mathcal{E} \), where variables $x_{in}$ and $x_{jm}$ is a binary that indicates function placement to node. The optimization problem can be defined as follows:

\begin{equation}
\begin{aligned}
    & \min_{x} \sum_{(i,j) \in \mathcal{E}} \sum_{n,m \in \mathcal{N}} L_{nm} x_{in} x_{jm} \\
    \text{s.t.} \quad & \sum_{i \in \mathcal{F}} D_{i} \leq R_{n} \quad \forall n \in \mathcal{N} \\
    & L_{nm} \leq S_{ij} \quad \forall (i,j) \in \mathcal{F}, \forall (n,m) \in \mathcal{N} \\
    & T_{\text{orb}}^{n}(t_i) + \sum_{i \in \mathcal{F}} T_{\text{exc}}^{in} \leq T_{\text{max}}^{n} \quad \forall n \in \mathcal{N} \\
    & x \in \{0,1\} \quad \forall i \in \mathcal{F}, \forall n \in \mathcal{N}
\end{aligned}
\end{equation}

The HyperDrive scheduling optimization model addresses key constraints of resource capacity, network SLOs, and temperature to guarantee efficient and reliable execution of Serverless workflows in the 3D Continuum.
Minimizing total latency while adhering to these constraints enables the scheduler to make placement decisions across diverse environments, ensuring optimal performance and system stability.
The consideration of satellite costs during scheduling is currently out of scope, since there are currently no pricing models for satellite nodes available.

\subsection{Heuristic Scheduling Algorithms for the 3D Continuum}
\label{sec:Scheduling.Algorithms}

Given the high computational complexity of the aforementioned optimization problem, heuristics are needed to allow implementing the HyperDrive scheduling model for the 3D Continuum.
We now examine the heuristic scheduling algorithms that approximate the aforementioned optimization problem.
To this end we rely on an \gls{MCDM} approach consisting of a sequence of filters that remove nodes that are not capable of hosting the task and scoring algorithms that determine the best suitable node among the eligible ones.

\subsubsection{Vicinity Selection}

Since the 3D Continuum may consist of tens of thousands of nodes, we need to perform a preselection of nodes before we can address the constraints of the optimization problem.
To this end, HyperDrive contacts the orchestrator to select a set of candidate nodes that are located in the vicinity of the desired location specified by the task or in the vicinity of its predecessor task.
The definition of the term ``vicinity'' can be configured independently for each part of the 3D Continuum.
For example, for the Cloud any data center node within a radius of 500~km of the desired location may be selected, while the radius could be 200~km for Edge nodes, and 2,000~km for satellites.
Akin to the vicinity, the total size of the candidates set and its composition can be configured as well, e.g., 500~total nodes consisting of 40\% Cloud nodes, 40\% Edge nodes, and 10\% Space nodes.

\subsubsection{Resource Checking}

After selecting the set of candidate nodes, HyperDrive first filters out all nodes that do not meet the resource requirements of the task.
Specifically, it checks the CPU architecture, CPU cores, memory, GPU (if present), local storage, and minimum battery charge (if the node has a battery) requested by the task.

\subsubsection{Network SLOs Enforcement}

HyperDrive uses a combination of filtering and scoring to ensure that the network \gls{QoS} \SLOs{} constraints for the incoming links of the task are fulfilled and the nodes with the best network properties are preferred.
For filtering we use \cref{alg:NetworkQoSPluginFilter}.
It iterates through all network \SLOs{} for incoming links, originating from predecessor tasks and external data sources (if any) and queries the network \gls{QoS} values for the lowest latency path between the candidate node and the node hosting the predecessor task or the data source.
If the network \SLO{} requirements are not met, the node is discarded.

For scoring we iterate through the aforementioned network paths again to determine the highest latency value
We assign the highest score, i.e., 100, to the node with the lowest latency and zero to the node with the highest latency; all nodes in between are assigned proportional scores in the target interval.

\begin{algorithm}[htb]
\caption{Network SLOs Filter.}
\label{alg:NetworkQoSPluginFilter}
\begin{algorithmic}[1]
    \Statex \textbf{Input: } $t$: Task to be scheduled;
    \Statex $cn$: Candidate node;
    \Statex $W = (V_W, E_W)$: Workflow DAG;
    \Statex $N = (V_N, E_N)$: Network graph;
    \Statex $S_t = \{(v, s) \forall v \in V_W\ s.t.\ (v, t) \in E_W \wedge s \neq \varnothing \}$: Network SLOs for incoming links of $t$;
    \Statex \textbf{Output: } $true$ if $cn$ can host $t$, otherwise $false$;
    
    \ForAll{$(v, s) \in S_t$}
        \State $u \gets$ \Call{GetHostNode}{$v, W, N$}
        \State $q \gets$ \Call{QueryNetworkQoS}{$u, cn, N$}
        \If{\Call{Latency}{$q$} $>$ \Call{MaxLatency}{$s$}}
            \State \Return $false$
        \EndIf
        \If{\Call{Bandwidth}{$q$} $<$ \Call{MinBandwidth}{$s$}}
            \State \Return $false$
        \EndIf
        \If{\Call{Jitter}{$q$} $>$ \Call{MaxJitter}{$s$}}
            \State \Return $false$
        \EndIf
        \If{\Call{PacketDrop}{$q$} $>$ \Call{MaxPacketDrop}{$s$}}
            \State \Return $false$
        \EndIf
    \EndFor
    \State \Return $true$
\end{algorithmic}
\end{algorithm}

\subsubsection{Temperature Optimization}

The algorithm to enforce the temperature constraint is geared specifically towards the Space part of the 3D continuum to prevent satellites from overheating due to excessive workload when in the sunlight.
Since a satellite that is close to overheating will reduce its computational power to prevent damage.
Thus, HyperDrive aims to prefer satellites, where the new task will not cause a problematic temperature.
This decision involves a complex estimate based on the current temperature of a satellite's compute unit, the expected duration of the task on the satellite's hardware, the required CPU and, possibly, GPU resources, the heat generated by these resources over the duration of the task, and the highest environmental temperature (based on in-orbit sunlight exposure) expected for the duration of the task.
This is encapsulated in the scoring logic of \cref{alg:TemperatureOptScore}.

\begin{algorithm}[htb]
\caption{Temperature Optimization Scoring.}
\label{alg:TemperatureOptScore}
\begin{algorithmic}[1]
    \Statex \textbf{Input: } $t$: Task to be scheduled;
    \Statex $cpu_t$: CPU cores requested by $t$;
    \Statex $gpu_t$: GPU cores requested by $t$;
    \Statex $n$: Node to be scored;
    \Statex $temp_{max}^n$: Maximum operating temperature for $n$;
    \Statex $temp_{rec}^n$: Recommended high temperature for $n$;
    \Statex \textbf{Output: } Score for the node $n$ in the range $[0; 100]$;
    
    \If{\Call{NodeType}{$n$} $\neq$ ``$satellite$''}
        \State \Return $100$
    \EndIf
    \State $d_t \gets$ \Call{GetExpectedDuration}{$t$}
    \If{$d_t == nil$}
        \Statex \Comment{If $d_t$ is unknown use the current temperature to compute the score.}
        \State $temp_{curr} \gets$ \Call{GetCurrTemp}{$n$}
        \State \Return \Call{CalcScore}{$temp_{curr}, temp_{rec}^n, temp_{max}^n$}
    \EndIf

    \Statex
    
    \State $temp_{inc} \gets$ \Call{EstimateCompTempIncrease}{$n, d_t, cpu_t, gpu_t$}
    \State $temp_{max}^{orb} \gets$ \Call{EstimateMaxOrbitTemp}{$n, d_t$}
    \State $temp_{max}^t \gets temp_{max}^{orb} + temp_{inc}$
    \State \Return \Call{CalcScore}{$temp_{max}^t, temp_{rec}^n, temp_{max}^n$}
    
    \Statex
    
    \Function{EstimateDuration}{$t$}
        \State $d_t \gets$ \Call{GetExpectedDuration}{$t$}
        \If{$d_t \neq nil$}
            \State \Return $d_t$
        \EndIf
        \State \Return \Call{MaxResponseTimeSLO}{$t$}
    \EndFunction

    \Statex \Comment{Estimates the temperature increase due to computation}
    \Function{EstimateCompTempIncrease}{$n, d_t, cpu_t, gpu_t$} 
        \State $temp_{inc} \gets$ \Call{CpuTempIncrease}{$n, cpu_t, d_t$}
        \State $temp_{inc} \gets temp_{inc}\ +$ \Call{GpuTempIncrease}{$n, gpu_t, d_t$}
        \State \Return $temp_{inc}$
    \EndFunction

    \Statex
    
    \Function{CalcScore}{$temp_{exp}, temp_{rec}, temp_{max}$}
        \If{$temp_{exp} \leq temp_{rec}$}
            \State \Return $100$
        \EndIf
        \If{$temp_{exp} > temp_{max}$}
            \State \Return $0$
        \EndIf
        \State $range \gets temp_{max} - temp_{rec}$
        \State $over_{rec} \gets temp_{exp} - temp_{rec}$
        \State \Return $\lfloor \left( 1 - \frac{over_{rec}}{range} \right) * 100 \rfloor$
    \EndFunction
\end{algorithmic}
\end{algorithm}

The algorithm first tries to get a duration estimate $d_t$ for the task.
This can be supplied by the user or through preceding profiling (on hardware similar to the satellite's) or the maximum response time \SLO{} of the task can be used.
If none of these values are available the score is calculated based on the current temperature of the satellite.
If $d_t$ value is available, it is used in conjunction with the requested resources to estimate the computation-based temperature increase $temp_{inc}$.
Subsequently, we determine the maximum expected environmental temperature $temp_{max}^{orb}$ during the orbit(s) within the duration of the task.
The sum of these two temperatures is the maximum expected temperature for the satellite during the execution of the task and is used for computing the node's score.
If the expected temperature is below the recommended temperature or above the maximum temperature, $100$ or zero are returned respectively.
Otherwise, a score is computed based on how much the temperature will go into the range between recommended and maximum temperature.

\subsubsection{Multi Commit}

Finally, all scores are accumulated for each node and the nodes are sorted by their scores.
The HyperDrive scheduler, then, contacts the orchestrator to assign the task to the highest scored available node using a multi-commit approach~\cite{VelaScheduler2023}.
Since multiple schedulers may be active, the orchestrator checks if the required resources are still available on the selected node.
If that is the case, the task is committed to the node, a success message is returned to the and the scheduler updates the information in the \gls{DAG} of the workflow instance.
If the orchestrator reports that the required resources are no longer available, the result is a scheduling conflict, which most distributed schedulers resolve by rerunning the scheduling pipeline.
To avoid doing this, HyperDrive tries committing the task to the second best node and, if that fails too, to the third best node, before triggering a rescheduling of the task.
This multi-commit technique has been shown to decrease the number of scheduling conflicts by a factor of 10 with respect to immediately rescheduling the task~\cite{VelaScheduler2023}.

\section{Implementation \& Experiments Design}
\label{sec:ImplementationAndExperimentsDesign}

To evaluate the HyperDrive scheduler we focus on the quality of the scheduling decisions and its scalability.
Since HyperDrive is, to the best of our knowledge, the first serverless scheduler specifically designed for the 3D Continuum, we compare it against three theoretical scheduling approaches: Greedy First-fit, Round-robin and Random scheduling.

\subsection{Implementation}

The prototype of the HyperDrive scheduler is implemented in Python as available as open-source\footnote{\HyperDriveSchedulerUrl{}}.
Since it is not feasible to run experiments on a \LEO{} satellite mega constellation, we have connected our scheduler to a modified version of the StarryNet satellite constellation simulator~\cite{StarryNet2023}.
The connection to the simulator is fully abstracted as an orchestrator interface, so that the simulator can be easily swapped.
StarryNet normally executes Docker containers for all nodes.
However, since we are interested in benchmarking the scheduling algorithms, we have replaced the containers with an in-memory nodes manager that tracks the available resources.

We have implemented the 3D Continuum-specific scheduling heuristics described in \cref{sec:Scheduling.Algorithms}.
StarryNet precomputes latencies between adjacent nodes for the entire duration of an experiment.
For each new time index, we use these latencies to update our network graph for network \SLOs{} enforcement.
Due to the absence of real satellite hardware information, we rely on reasonable estimates for the temperature optimizations.

\subsection{Experiments Design}

With our experiments we evaluate two critical aspects of the HyperDrive scheduler: 
\begin{enumerate*} [(i)]
    \item scheduling quality with respect to latency and satellite temperature management and
    \item scalability.
\end{enumerate*}

For assessing the scheduling quality we examine two major quality objectives.
The primary objective is the latency achieved between the individual tasks of a serverless workflow and the \gls{E2E} latency.
The secondary objective is the intelligent selection of satellite nodes with respect to their temperature situation, i.e., satellites should be chosen, which will not overheat and reduce computational power while processing a task.

\begin{table}
\centering
\caption{Infrastructure Sizes used for Evaluation.}
\label{tab:InfrastructureSizes}
\begin{tabular}{|r|r|r|r|}
\hline
\multicolumn{1}{|c|}{\textbf{Satellites}} &
  \multicolumn{1}{c|}{\textbf{Edge Nodes}} &
  \multicolumn{1}{c|}{\textbf{Cloud Nodes}} &
  \multicolumn{1}{c|}{\textbf{Total Nodes}} \\ \hline
1,008 & 100 & 10 & \textbf{1,118} \\ \hline
2,016 & 200 & 20 & \textbf{2,236} \\ \hline
3,024 & 300 & 30 & \textbf{3,354} \\ \hline
4,032 & 400 & 40 & \textbf{4,472} \\ \hline
\end{tabular}
\end{table}

To set up the experiment we use TLE data, obtained on July 2, 2024 from CelesTrack\footnote{\url{https://celestrak.org/NORAD/elements/}}, describing the orbits of \StarlinkNodesCount{} nodes of the Starlink\footnote{\url{https://www.starlink.com}} LEO satellite constellation.
We deploy our wildfire detection use case, whose workflow is shown in \cref{fig:workflow}.
We assign the \inlinecode{Ingest} function to a drone flying over a region of California, USA that is prone to wildfires and trigger the scheduling of the remaining functions as the simulation progresses.
Since the StarryNet only supports satellite and ground station nodes, we model the drone as a ground station node.
Since we evaluate the scheduling at the time when the second function needs to be placed, we do not require any movement from the drone, hence modeling it as a ground station does not limit our evaluation scenario.
All experiments are run using Python~3.12 on Ubuntu~20.04 LTS on a Windows Subsystem for Linux~2 VM with 8~vCPUs and 8~GB of RAM.
The VM is hosted on a laptop running Windows~10 22H2 on a Whiskey Lake-U generation Intel Core i7 processor.

We benchmark HyperDrive against the following theoretical schedulers, which we use as baselines:
\begin{itemize}
    \item Greedy First-fit
    \item Round-robin
    \item Random selection
\end{itemize}


For evaluating the scalability we want to examine how HyperDrive scales with respect to the infrastructure size.
To this end, we benchmark the placement of wildfire detection workflow on increasing infrastructure sizes.
For Cloud and Edge nodes we simulate nodes in the region the workflow is deployed in, while for satellites we simulate an entire constellation with the current 72 orbital planes of Starlink and an equal number of satellites per plane.
Specifically, we use the infrastructure sizes described in \cref{tab:InfrastructureSizes} -- node that the numbers in this table refer to our simulation only, which is limited by the resources of our host machine.

We execute five iterations of every scheduler's placement of the wildfire detection workflow on each of the four infrastructure sizes.
We examine the achieved \gls{E2E} latencies and temperature characteristics to evaluate the scheduling quality of all four schedulers and HyperDrive's processing time per task to assess its scalability.

\section{Experimental Results}
\label{sec:Results}

\subsection{Scheduling Quality}

To evaluate the scheduling quality we examine the network latencies achieved by the placements and the temperatures of the selected satellites (if any).

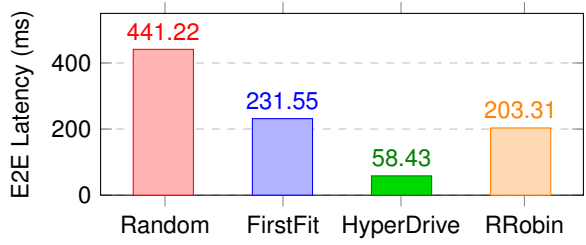
\begin{figure}[t]
    \centering
    \begin{tikzpicture}
        \begin{axis}[
            ybar,
            width=8cm,
            height=4cm,
            xlabel={},
            ylabel={E2E Latency (ms)},
            xtick={-0.8, 1.4, 3.6, 5.9},
            xticklabels={Random, FirstFit, HyperDrive, RRobin},
            nodes near coords,
            nodes near coords style={font=\sansmath\sffamily\small},
            ymin=0,
            ymax=550,
            bar width=0.8cm,
            ymajorgrids=true,
            grid style=dashed,
            enlarge x limits=1
        ]
        \addplot[color=red, fill=red!30] coordinates {(1, 441.22)};
        \addplot[color=blue, fill=blue!30] coordinates {(2, 231.55)};
        \addplot[color=green!50!black, fill=green!85!black] coordinates {(3, 58.43)};
        \addplot[color=orange, fill=orange!30] coordinates {(4, 203.31)};
        \end{axis}
    \end{tikzpicture}
    \caption{Wildfire Detection Workflow Mean E2E Latency per Scheduler.}
    \label{fig:E2E_Latency}
\end{figure}
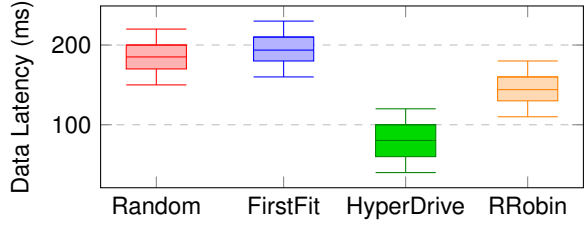
\begin{figure}
    \centering
    \begin{tikzpicture}
        \begin{axis} [
            box plot width=4mm,
            width=8cm,
            height=4cm,
            ylabel={Data Latency (ms)},
            xtick={-1, 1.4, 3.7, 6},
            xticklabels={Random, FirstFit, HyperDrive, RRobin},
            ymajorgrids=true,
            grid style=dashed,
            enlarge x limits=0.15
        ]
        \boxplot [red, fill=red!30] {EODataLatency.dat}
        \boxplot [color=blue, fill=blue!30] {EODataLatency2.dat}
        \boxplot [green!50!black, fill=green!85!black] {EODataLatency3.dat}
        \boxplot [orange, fill=orange!30] {EODataLatency4.dat}
        \end{axis}
    \end{tikzpicture}
    \caption{Data Latency per Scheduler}
    \label{fig:EO_DataLatency}
\end{figure}

\cref{fig:E2E_Latency} shows the mean network \gls{E2E} latencies achieved by the four schedulers across all 20~experiment iterations, i.e., five iterations for each of the four infrastructure sizes.
For clarity, the shown latencies are the sum of the network latencies only, without function execution times.
The \gls{E2E} network latency \SLO{}, without function execution times, across all four functions of the wildfire workflow is 350~ms.
While all schedulers, except for the Random scheduler, meet the \gls{E2E} network latency \SLO{}, HyperDrive clearly has the lowest latency, because it actively optimizes for it.
HyperDrive's \gls{E2E} latency is 71\% lower than Round-robin's, which is the second best.
While Greedy First-fit and Round-robin meet the \gls{E2E} network \SLO{}, they violate individual function network \SLOs{} in about 33\% of the cases for Greedy First-fit and in 30\% of the cases for Round-robin.
HyperDrive fulfills all function network \SLOs{}.

Apart from inter-function network \SLOs{}, the wildfire detection workflow also defines a network \SLO{} for an EO satellite data source.
The \inlinecode{object-det} function requires a maximum latency of 175~ms to the respective EO satellite.
\cref{fig:EO_DataLatency} shows the EO data latencies achieved by the schedulers.
Random and Greedy First-fit violate the \SLO{}.
HyperDrive and Round-robin fulfill it on average, albeit Round-robin violates the \SLO{} in 35\% of the cases.
HyperDrive always fulfills it, because its filtering does not allow scheduling on nodes that would violate the \SLO{}.

\begin{figure}[t]
  \hspace{2.7em}
  \begin{tikzpicture}
      \begin{axis}[
      mesh/cols=3, 
      mesh/rows=3,
      mesh/check=false,
      enlargelimits=false,
      colorbar,
      colormap/Reds,
      height=6cm,
      width=7.2cm,
      xtick={0, 1, 2},
      xticklabels={HyperDrive,Random, RRobin},
      ytick={0, 1, 2},
      yticklabels={object-det, prepare-ds, extract-frames},
      yticklabel style={rotate=90,font=\sansmath\sffamily\scriptsize},
      nodes near coords,
      nodes near coords style={font=\sansmath\sffamily\small},
      point meta=explicit symbolic,
      nodes near coords align={center},
      colorbar style={
            title={Overheating Temperature (°C)},
            yticklabel style={font=\sansmath\sffamily\footnotesize},
            ytick={0, 4, 8, 12, 16},
            title style={rotate=90,font=\sansmath\sffamily\footnotesize},
            at={(1.05,0.5)}, 
            title style={yshift=-1.15cm,xshift=-2.5cm},
            anchor=west
        }
      ]
      \addplot [matrix plot,point meta=explicit]
          coordinates {
            (0,0) [0] (1,0) [14.0] (2,0) [14.0]
    
            (0,1) [0] (1,1) [4.85] (2,1) [4.85]
    
            (0,2) [0] (1,2) [2.90] (2,2) [2.90]
          };
         \addplot [matrix plot,point meta=explicit,text=white]
          coordinates {
            (1,0) [14.0] (2,0) [14.0]
          };
      
      \end{axis}
    \end{tikzpicture}
    \caption{Scheduling Overheating Map.}
    \label{fig:Overheating}
\end{figure}
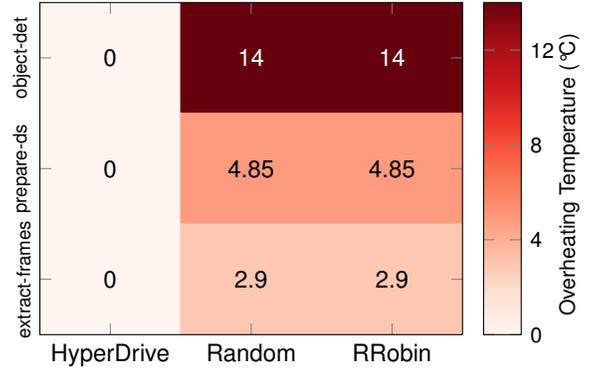

The secondary optimization objective after the network latency, is satellite temperature measurement to avoid overheating.
\cref{fig:Overheating} shows a heat map for the three scheduled functions that documents cases when the functions are scheduled on satellites and their temperature exceeds the recommended operating temperature.
HyperDrive places 34 of the total 60 function instances (56.7\%) across all iterations on satellites and never exceeds the recommended temperature.
The Random scheduler places 56 of 60 function instances (93.3\%) on satellites and exceeds the recommended temperature in 23 (41\%) of these cases; in three cases it even exceeds the maximum operating temperature.
Round-robin schedules all 60 function instances on satellites and exceeds the recommended temperature in one third of the cases; in four cases it exceeds the maximum operating temperature.
It should be noted that as the number of Edge nodes increased in the two larger infrastructure sizes, HyperDrive selected more Edge nodes instead of satellites, due to their favorable network latencies; for the smaller two infrastructure sizes 86.7\% of the nodes were satellites, while for the larger two only 33.3\% were satellites.

\subsection{Scalability}

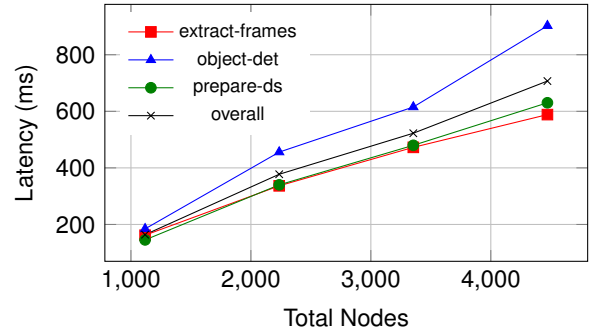
\begin{figure}
    \hspace{0.2em}
    \begin{tikzpicture}
        \begin{axis}[
            xlabel={Total Nodes},
            ylabel={Latency (ms)},
            legend style={draw=none, font=\sansmath\sffamily\scriptsize},
            legend pos=north west,
            grid=major,
            height=5cm,
            width=8cm
        ]
        \addplot[color=red, mark=square*] coordinates {
            (1118, 162.00)
            (2236, 336.60)
            (3354, 472.60)
            (4472, 588.40)
        };
        \addlegendentry{extract-frames}
        \addplot[color=blue, mark=triangle*] coordinates {
            (1118, 184.40)
            (2236, 455.80)
            (3354, 615.20)
            (4472, 902.40)
        };
        \addlegendentry{object-det}
        \addplot[color=green!50!black, mark=*] coordinates {
            (1118, 145.40)
            (2236, 340.00)
            (3354, 479.80)
            (4472, 630.20)
        };
        \addlegendentry{prepare-ds}
        \addplot[color=black, mark=x] coordinates {
            (1118, 163.93)
            (2236, 377.47)
            (3354, 522.53)
            (4472, 707.00)
        };
        \addlegendentry{overall}
        \end{axis}
    \end{tikzpicture}
    \caption{HyperDrive Scheduling Latency Across Infrastructure Sizes.}
    \label{fig:Scalability}
\end{figure}

The goal of the scalability evaluation is to see how HyperDrive's performance evolves as the infrastructure size increases.
\cref{fig:Scalability} shows the mean scheduling latency for each of the three serverless functions as well as the overall average.
Since the prototype implementation is not connected to a real orchestrator and manually performs the vicinity selection with a linear search, we disregard the nominal scheduling latency values and focus on how they evolve with increasing infrastructure sizes.

It is evident that HyperDrive's performance scales linearly with the infrastructure size.
The \inlinecode{object-det} function has a steeper incline than the others or the overall average, because needs to check twice as many network \SLOs{} as the others, because it has a data source network \SLO{}.
Nevertheless, its increase remains linear.

\subsection{Discussion}

As previously seen, HyperDrive is the only scheduler specifically designed for the challenges of the 3D Continuum.
HyperDrive excels at choosing between satellite and terrestrial nodes, depending on what benefits the network \SLOs{} the most.
As more nodes are available the quality of its scheduling decisions improves, e.g., the mean network latency between functions drops by 73\% in the larger two infrastructures compared to the smaller two infrastructures.

Larger infrastructures yield better scheduling results, but they also increase processing time.
Increased processing time, however, does not offset the benefits of optimized scheduling, because transfer and processing times of EO data are orders of magnitude greater than the scheduling duration.
Additionally, scheduling on LEO satellites typically does not require high scheduling throughput, due to the type of applications that are expected to be deployed, e.g., federated learning in space or at the Edge~\cite{FLinSatelliteConstellations2024, gajanin2024HARFederatedLearning}, advanced automotive use cases~\cite{IntelligentSatellitePayloads6G2024}, monitoring applications~\cite{SECO2024}, or disaster relief~\cite{FireSatBlog2024}.

Finding multiple shortest paths through a large network graph is the biggest concern to the performance of HyperDrive.
While HyperDrive scales linearly with the infrastructure size, the path finding time can be reduced by using a hypergraph to reduce the number of links and by computing paths between regions instead of single nodes.
Additionally, the paths can be periodically precomputed and cached by the orchestrator.
This will be addressed by our future work.

Currently we assume the absence of congestion on the network routes, but as satellite usage increases, this will be considered in future work. Additionally, a dense constellation can provide multiple routes~\cite{DelayNotAnOption2018} between two nodes and prioritization can be employed for disaster response applications.

We evaluated HyperDrive in simulations.
However, its scheduling algorithms can be transitioned to a physical system.
To this end they must be connected to a real-world orchestrator, which supplies metadata about real satellites (as well as Edge and Cloud nodes) and which can deploy functions on these nodes.

\section{Related Work}
\label{sec:Relatedwork}

We now discuss other work that is related to ours and compare HyperDrive to it. 

\subsection{Edge Cloud Continuum \& Orbital Edge Computing}

Several research studies~\cite{SlocVision2020,Dustdar2021,EdgeResearchOpportunities2023,DCCS_2022,SCF_2022} have proposed a paradigm known as the Edge-Cloud continuum (ECC). This paradigm involves integrating computing resources in different layers, composed of Edge devices such as sensors and wearables that produce data processed by low-resource Edge nodes close that are close to the Edge devices and high-resource Cloud servers. ECC aims to enable seamless integration between all the layers, allowing for efficient task distribution and improved application performance. By utilizing the advantages of both Edge and Cloud resources, ECC enables heterogeneous environments to adjust to computational needs and connectivity conditions. In this paper, HyperDrive proposes to expand the ECC to orbit by seamlessly integrating satellites as Edge nodes, thus creating a 3D Edge-Cloud-Space Continuum.

Lately, the extensive effort to expand on-orbit capability lead further research to explore the implementation of core networks in space, offering several benefits such as enhancing mobile coverage in remote areas, facilitating direct device-satellite connections, and satellite computing~\cite{Empowering6G2023, Evolution5Gto6G2022,Earth2Space2023,L2D2_2021}. 

LEO satellites, like terrestrial Edge nodes, have limited computing capacity and like Edge nodes, satellites can be near data sources, such as Earth observation satellites. Therefore, the increase in LEO satellites in orbit allows data to be processed directly in orbit, near the data source, enabling Orbital Edge Computing (OEC)~\cite{OECNano2020,InorbitExperiment2020}. Research~\cite{FLOEC2024,SatelliteBasedFL2022,chen2023edge,optimizingFLscheduling2023,Fool2024} enables federated learning by leveraging their distributed localized data processing capabilities, enhancing real-time data analysis and decision-making in space applications.

The Tiansuan~\cite{SATVisionAndChallenges2023,SCCaseStudy2023} constellation leverages a cloud-native design to enhance onboard services, resources, and the development and management of satellite equipment. Tiansuan's cloud-native approach provides advantages in application deployment, scalability, and cost-effectiveness compared to traditional satellite designs, allowing for seamless integration of computing, and networking. Tisuan's platform is composed of six different layers: Physical, Virtual Resource, Operating System, Container Service, Collaborative Orchestration, and Function  Application.
MobileViT~\cite{SatIoTSmartAgri20223} propose a three-layer architecture to enable Satellite Internet of Things for Smart Agriculture. The infrastructure layer is composed of IoT devices such as sensors, drones and satellites. The capacity layer contains computing communication and caching while the application layer represents the different use cases such as Smart Agriculture, Smart Grid and Smart Port.

\subsection{Space-as-a-Service}

Research identifies emerging services in space~\cite{SatelliteConstellations2021,TaxonomonySpaceAsAService2022} such as 
\emph{Constellation-as-a-Service}, \emph{Satellite-as-a-Service} and \emph{Payload-as-a-Service}. 

\paragraph{Constellation-as-a-Service} Mission MP42~\cite{NanoAvionics2024} by NanoAvionics and Satellogic~\cite{ConstellationService2024} aims to offer a satellite constellation service to IoT/M2M operators. Constellation-as-a-service allows businesses to deploy and manage their satellite network without launching their own spacecraft. This business model promises customized services such as dedicated satellites, dedicated rocket launches, in-country operation centers, access to a global ground station network, and dedicated platforms, including a private Cloud for image cataloging, processing, and storage.

\paragraph{Satellite-as-a-Service} It proposes a shared multi-tenant satellite concept~\cite{Orbitals_Whitepaper2024,QoEAwareSatas2024,SatasS2020}. The shared-access model allows multiple missions to be hosted on a single satellite, enabling users to share platform and payload capabilities. The Satellite as a Service model includes ground segment validation using continuous integration and hardware simulators to ensure the reliability and safety of user-uploaded software. This approach uses existing satellite infrastructure and modern software tools such as CI/CD to create a flexible and cost-effective platform for space technology development.

\paragraph{Payload-as-a-Service} It emerged after a shift from analog to digital satellite payload, that enabled satellites to serve multiple clients. Digital payload made it possible to customize satellite computing for specific purposes such as machine learning and earth observation~\cite{GMV_Evolution2024}. Consequently, it enabled an alternative to expensive space infrastructure - \emph{Payload-as-a-Service}. In this business model, a commercial operator owns and manages the satellite system, providing data (payload) to customers on demand. The service providers manage the satellite bus, integration, launch, and operations. On the other hand, clients access the data and may even operate the payload by starting/stopping data collection and monitoring.~\cite{TaxonomonySpaceAsAService2022,senior2021can,SustainableSAT2022}.

All of these approaches focus mostly on satellites only, with very little or no involvement of terrestrial compute nodes.
HyperDrive, proposes an unified computing continuum that spans seamlessly across Edge, Cloud, and Space nodes.
As such HyperDrive goes further than the aforementioned concepts.
But the HyperDrive scheduler can also complement the Constellation-as-a-Service and Satellite-as-a-Service, because both allow customers to run their own workloads on satellites and, thus, require a scheduling mechanism.

\subsection{Satellite Edge Task Scheduling \& Offloading} 

In~\cite{DNN2023}, an efficient framework is proposed for offloading inference tasks by partitioning Deep Neural Network (DNN) models into multiple satellites, including one high Earth orbit (HEO) satellite and multiple LEO satellites. The approach divides inference tasks, with the task owner executing the initial portion of the DNN and offloading the remaining portion to other satellites. FedLEO~\cite{optimizingFLscheduling2023} proposes a distributed scheduling mechanism for LEO satellite constellations to overcome bandwidth limitations and intermittent connectivity. FedLEO leverages Satellite Edge Computing (SEC) to improve training efficiency by adding horizontal communication pathways among satellites and optimally scheduling interactions with ground stations. Unlike HyperDrive, FedLEO and task offloading approach considers data processing only satellites SEC, thus excluding Edge nodes in the ground for task placement. 

An Orbital Edge (OE)~\cite{OrbitaEdgeOffloading2022} platform leverages ISL for satellite processing, reducing latency and leveraging distributed computational capabilities. It offloads computing tasks from ground nodes to a single satellite and its neighboring satellites. While the OE platform relies on satellite-ground communication links, which may cause data transfer delays, HyperDrive addresses each computing layer's challenges separately to create a unified Edge Cloud and Space Continuum platform.


\section{Conclusion}
\label{sec:Conclusion}

We presented HyperDrive, a novel Serverless platform that is specifically designed to enable a seamless execution of Serverless workflows across the 3D Continuum.
We discussed the unique challenges of the 3D Continuum, such as the short availability windows of the fast moving LEO satellites, solar power supply, and the possibility of overheating while facing the sun.
The HyperDrive scheduler enables the optimized placement of Serverless functions in the 3D Continuum by considering network \SLOs{}, workflow and data source dependencies, and thermal conditions of satellites.
HyperDrive is, to the best of our knowledge, the first Serverless scheduler for the 3D Continuum.
We evaluate it against three theoretical baseline schedulers by scheduling a wildfire disaster response workflow with strict network \SLOs{} and EO satellite data dependencies.
HyperDrive achieves 71\% lower \gls{E2E} network latency than the best baseline and shows linear performance scalability with the infrastructure size.

As future work we plan to continue our realization of the HyperDrive Serverless platform for the 3D Continuum.
An important goal is to further to improve the coordination of function execution and satellite orbits, by placing functions on satellites that will be in the ideal position for a low-latency handoff to the next node when the function completes.
We also intend to reduce scheduling complexity for large infrastructure sizes by using hypergraphs for inter-node path computations.
Additionally, we envision a lightweight framework for Serverless-native development of the next generation EO applications for the 3D Continuum.

\balance

\ifthenelse{\boolean{anonymousMode}}{}{
    \section*{Acknowledgment}
        This work is partially funded by the Austrian Research Promotion Agency (FFG) under the project RapidREC (Project No. 903884).
        
        This research received funding from the EU’s Horizon Europe Research and Innovation Program under Grant Agreement No. 101070186. EU website for TEADAL: \url{https://teadal.eu}.

}

\bibliographystyle{IEEEtran}
\bibliography{IEEEabrv, bibliography}

\end{document}